\documentclass[%
reprint,
superscriptaddress,
 amsmath,amssymb,
 aps,
prl
]{revtex4-2}

\usepackage{graphicx}%
\usepackage{dcolumn}%
\usepackage{bm}%
\usepackage{hyperref}%

\usepackage{xcolor}   %

\begin{document}

\title{Efficient and versatile model for vibrational STEM-EELS}%

\author{Paul M. Zeiger}
\email{paul.zeiger@physics.uu.se}
\affiliation{%
  Department of Physics and Astronomy, Uppsala University, P.O. Box 516, 75120 Uppsala, Sweden
}%
\author{J\'{a}n Rusz}%
\affiliation{%
  Department of Physics and Astronomy, Uppsala University, P.O. Box 516, 75120 Uppsala, Sweden
}%

\date{\today}%

\begin{abstract}
We introduce a novel method for the simulation of the impact scattering in vibrational scanning transmission electron microscopy electron energy loss spectroscopy (STEM-EELS) simulations. The phonon-loss process is modeled by a combination of molecular dynamics and elastic multislice calculations within a modified frozen phonon approximation. The key idea is thereby to use a so-called $\delta$-thermostat in the classical molecular dynamics simulation to generate frequency dependent configurations of the vibrating specimen's atomic structure. The method includes correlated motion of atoms and provides vibrational spectrum images at the cost comparable to standard frozen phonon calculations. We demonstrate good agreement of our method with simulations and experiments for a 15~nm flake of hexagonal boron-nitride (hBN).
\end{abstract}

\maketitle

Excellent spatial resolution is the highlight of the transmission electron microscope \cite{haid+98nat392,bats+02nat418,akas+15apl106,sawa+15microscopy64,mori+16apl108,mori+18micromicroanal24}, allowing atomically resolved elemental mapping and chemistry. Recent instrumental developments have improved the energy resolution of electron energy loss spectroscopy down to 4.2~meV \cite{kriv+14nat,kriv+19ultramic203}. This unique combination of high spatial and energy resolution opened doors to unprecedented experiments such as temperature measurement at the nano-scale \cite{idro+18prl120,lago-bats18nanolett18}, identification and mapping of isotopically labeled molecules \cite{hach+19sci363}, position- and momentum-resolved mapping of phonon modes \cite{hage+18sciadv4,seng+19nat}, mapping of bulk and surface modes of nanocubes \cite{lago+17nat543}, or investigations of the nature of polariton modes in van der Waals crystals \cite{govy+17natcomm8}.

Inelastic electron scattering on atomic vibrations in a polar material consists of two major contributions, namely, impact scattering and dipolar scattering \cite{dwye14prb89,dwye+16prl117}. The latter stems from a long ranged interaction between the beam electron and an oscillating dipole moment generated by the atomic vibrations. It can be used to perform damage-free vibrational spectroscopy in an aloof beam geometry \cite{croz+16ultramic169,rez+16ncomms7}. High-angle impact scattering on the other hand is quite localized scattering on the atomic potential and allows for high resolution imaging and spectroscopy \cite{egoa+14ultramic147,dwye+16prl117,hage+19prl,venk+18arXiv1812.08895}.

From a theoretical point of view, there is a limited number of approaches one can follow to simulate vibrational spectroscopy and imaging. Dwyer presents a treatment of both single impact and single dipolar scattering including a detailed phonon dispersion from a density functional theory calculation in a multislice scheme \cite{dwye17prb96}. Forbes et al.\ introduced a quantum excitation of phonons model, which allows for the inclusion of thermal diffuse scattering into multislice image simulations \cite{forb+10prb}. The thermal vibrations are modeled within the Einstein model for quick image calculations, neglecting correlations in atomic movements and dispersion effects. Such an approach is not able to simulate vibrational spectra but in principle one could use different models of atomic vibrations to generate the necessary configurations. In a more recent work, Forbes et al.\ described a method to model the vibrational sector of the electron energy loss spectrum \cite{forb-alle16prb}. Their method is based on an explicit evaluation of inelastic matrix elements, but neglects the strong elastic interaction of electrons with the specimen before and after the inelastic event. Lastly Nicholls et al.\ presented a method to simulate momentum-resolved vibrational electron energy based on the Van Hove scattering formalism and carried out using density functional theory, but neglecting dynamical diffraction \cite{nich+19prb99}.

Further deployment of vibrational spectroscopy at high spatial resolution calls for an efficient and accurate computational tool capable to predict or interpret the experiments under realistic conditions. The method should allow simulations of temperature-dependent vibrational spectrum images of arbitrary systems, including interfaces, impurities or defects to interpret minute changes in vibrational spectra as a function of scattering angles.

In this Letter, we present a method for simulating vibrational electron energy loss spectra including dynamical diffraction effects. %
The method is based on molecular dynamics calculations and the frozen phonon approximation, from which it inherits its computational efficiency.
It can be applied to arbitrary systems, for which suitable interatomic potentials exist. It is computationally efficient, versatile and conceptually transparent.

\begin{figure}
    \includegraphics[width=\linewidth]{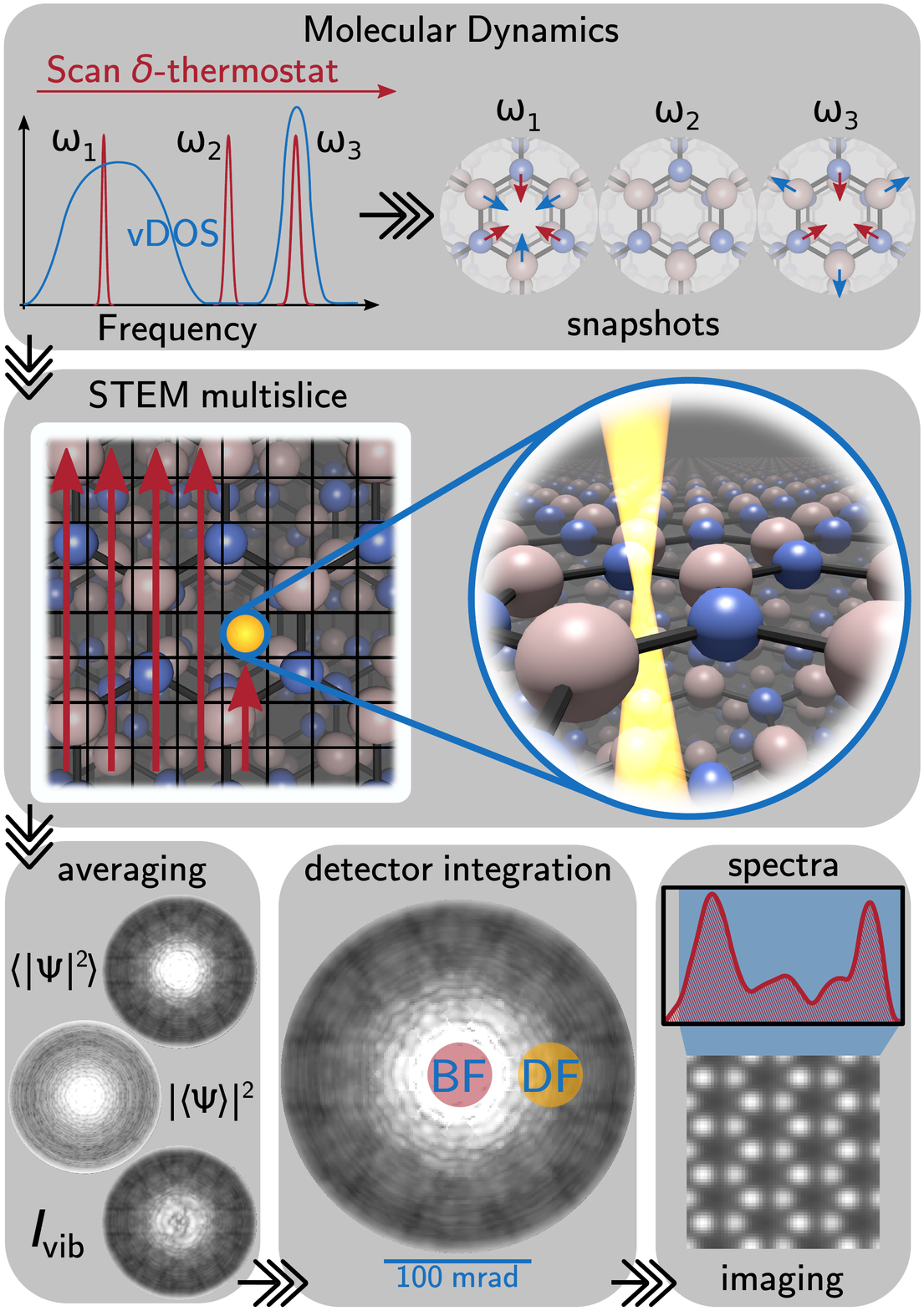}
    \caption{Schematic overview of our simulation method. The idea is to generate sets of snapshots for each energy bin, which is excited by the $\delta$-thermostat. The snapshots are subsequently used to simulate the propagation of an elastic electron wave function through the crystal using the multislice method. Thereafter the difference $\langle |\Psi|^2 \rangle - | \langle \Psi\rangle|^2$ of incoherent and coherent snapshot averages of the exit wave function $\Psi$ gives access to the vibrationally scattered intensity $I_{\mathrm{vib}}$. Integration over detector collection angles yields spectra, which can be used to form images of the specimen. Refer to the main text for more details on the calculation process and used parameters.}
    \label{fig:method_schematics}
\end{figure}

The frozen phonon model is a semi-classically motivated approximation, in which vibrational effects are taken into account by suitably averaging over "frozen" configurations of the vibrating atomic structure called "snapshots" \cite{loan+91actacrystA47}. Quantum mechanical considerations have established the validity of its approximations \cite{wang1995book,wang98actacryst,dyck09ultramic}. The frozen phonon model has been successfully used to account for thermal diffuse scattering, either in conjunction with the Einstein model of atomic vibrations, for which it delivers numerically equivalent results to the quantum excitations of phonons model \cite{forb+10prb}, or with models including a detailed phonon dispersion \cite{mobu+98actacrystA54,mull+01ultramic86,bisk+07micromicroanal13,avey+14prl113,lofg+16ultramic164,pohl+17scirep7,krau+18ultramic189}.

Classically, thermostats are used in Molecular Dynamics simulations to enforce thermodynamical conditions, but recent research on stochastic thermostats based on the non-Markovian generalized Langevin equation has shown, that such thermostats can be tailored to a wide range of applications \cite{ceri+10jctc}, among others, the efficient thermostatting of path integral molecular dynamics simulations \cite{ceri+10jcp133}, the simulation of nuclear quantum effects in solids at low cost \cite{ceri+09prl103}, and the frequency dependent heating of normal modes, the so-called $\delta$-thermostat \cite{ceri-parr10pcs}. This thermostat artificially heats those vibrational modes, whose frequencies lie within a narrow range of frequencies $\Delta \omega$ around a chosen peak frequency $\omega_0$. The thermostat enforces canonical sampling at $\omega_0$ at a given temperature $T^\star$. Modes, whose frequencies lie outside of $\Delta \omega$, are kept at a much lower temperature, effectively "freezing" them. In the context of electron microscopy, nuclear quantum effects modeled by a colored noise thermostat have readily been included into simulations of temperature-dependent thermal diffuse scattering in high angle annular dark field images \cite{lofg+16ultramic164}.

Our method combines Molecular Dynamics simulations using a $\delta$-thermostat and multislice simulations to a frequency-resolved frozen phonon method. The key idea is to scan the peak frequency $\omega_0$ of the $\delta$-thermostat over the range of vibrational frequencies of the studied material and at each energy bin $\omega_i$, a set of $N$ snapshots $\left\lbrace \bm{\tau}_n \right\rbrace$ is generated. Subsequently an exit wave function $\Psi\left(\bm{k}_{\perp}, \bm{r}_{\textrm{b}}, \bm{\tau}_n(\omega)\right)$ is computed for each snapshot and each STEM beam position $\bm{r_{b}}$ at a given thickness using the multislice method. This process is iterated over all energy bins and the vibrational energy loss spectrum is successively assembled in this way.

In order to separate the vibrationally scattered electron intensity $I_{\mathrm{vib}}(\bm{k}_{\perp}, \bm{r}_{\textrm{b}}, \omega)$, the electron intensity is averaged incoherently and coherently over the number of snapshots $N$ in the spirit of the quantum excitations of phonons model \cite{forb+10prb}. The difference between both averages is the vibrationally scattered intensity.
We have thus:
\begin{eqnarray}
    I_{\mathrm{vib}}(\bm{k}_{\perp}, \bm{r}_{\mathrm{b}}, \omega) = & {} \frac{1}{N} \sum_{n=1}^{N} \left|\Psi\left(\bm{k}_{\perp}, \bm{r}_{\mathrm{b}}, \bm{\tau}_n(\omega) \right) \right|^2 \nonumber\\
    & {} -  \frac{1}{N^2} \left| \sum_{n=1}^{N} \Psi\left(\bm{k}_{\perp}, \bm{r}_{\mathrm{b}}, \bm{\tau}_n(\omega)\right) \right|^2. \label{eq:Ivib_avg}
\end{eqnarray}
The vibrational intensity is integrated over the detector area $\Omega$ to yield spectral information at each beam position $\bm{r}_{\mathrm{b}}$, i.e.,
\begin{equation}
    I_{\mathrm{vib}}(\bm{r}_{\mathrm{b}}, \Delta E) = \int_{\Omega} \mathrm{d}\bm{k}_{\perp} \; I_{\mathrm{vib}}(\bm{k}_{\perp}, z, \bm{r}_{\mathrm{b}}, \Delta E). \label{eq:Ivib_spectra}
\end{equation}
Images are formed by integrating the spectra over a specific energy window at every beam position, i.e.,
\begin{equation}
    I_{\mathrm{vib}}^{E_-,E_+}(\bm{r}_{\mathrm{b}}) = \int_{E_-}^{E_+} \mathrm{d} \Delta E \; I_{\mathrm{vib}}(\bm{r}_{\mathrm{b}}, \Delta E). \label{eq:Ivib_image}
\end{equation}
The obtainable energy resolution of our method is approximately given by the range of heated frequencies $\Delta \omega$. Currently $\Delta \omega$ is implemented as a relative resolution $\Delta \omega = 0.01 \cdot \omega$. For vibrational modes of energies between 50-200~meV, this frequency resolution translates to an energy resolution of 0.5-2~meV.

In order to put our method to the test we chose simulation parameters, which follow the experiment and simulation of Ref.~\cite{hage+19prl}. The simulation box contains 46 layers of hBN in AA' stacking order, each of which consists of 448 atoms in $8\times14$ conventional cells. Periodic boundary conditions apply in all three spatial directions and the dimensions of the relaxed simulation box are 34.59~{\AA} $\times$ 34.95~{\AA} $\times$ 148.13~{\AA}.

MD simulations are carried out using a combination of the i-Pi and LAMMPS software packages \cite{ceri+14cpc185,kapi+18cpc,plim95jcompphys117,lammpsweb}. At each time step LAMMPS is used to evaluate the forces on the nuclei using an empirical potential and i-PI performs the time integration and thermostatting. The intra and inter layer interactions of the nuclei are described by a Tersoff and a specifically for hBN developed inter layer potential \cite{sevi+11prb,ouya+18nl}. Additionally a shielded Coulomb potential accounts for partial charges as in Ref.~\cite{maar+17jpcc}. All these potentials are readily implemented in LAMMPS.

$n_E = 18$ different molecular dynamics trajectories are simulated, one for each energy bin in the range between 11.5 to 54.0~THz with an increment of 2.5~THz, which correspond to phonon energies of around 50 to 220~meV. The peak temperature $T^\star$ of the $\delta$-thermostat is set to 300~K and the modes outside the heated range of frequencies are maintained at $10^{-4}\cdot T^\star$. The GLE matrices defining the properties of the thermostat are obtained from an online repository \cite{ceri+09prl102,gle4md_web}. The structure and simulation box size are relaxed using the conjugate-gradient method as implemented in LAMMPS at a target pressure of 0~bar before starting the time integration. A time step of 0.5~fs is used to simulate the molecular dynamics trajectories and snapshots are taken every 1900--2000 time steps after an initial equilibration of 50000 time steps. In this way $n_{\textrm{snap}} = 64$ snapshots are generated for each energy bin.

Elastic multislice calculations using the real space multislice method \cite{cai+09micron} are performed on a numerical grid of $N_x \times N_y \times N_z = 672 \times 672 \times 1532$ points. The STEM grid was chosen such that one conventional cell of hBN is covered with $28\times 16$ beam positions, of which only one quarter needs to be calculated thanks to its symmetry. The electron beam is simulated for an acceleration voltage of 60~keV, convergence angle of 31.5~mrad, and incident along the [0001] zone axis of hBN. In order to account for the finite energy resolution of real microscopes the spectra are broadened by a rectangular function of 20~meV width and unit integral.

\begin{figure}
    \includegraphics[width=\linewidth]{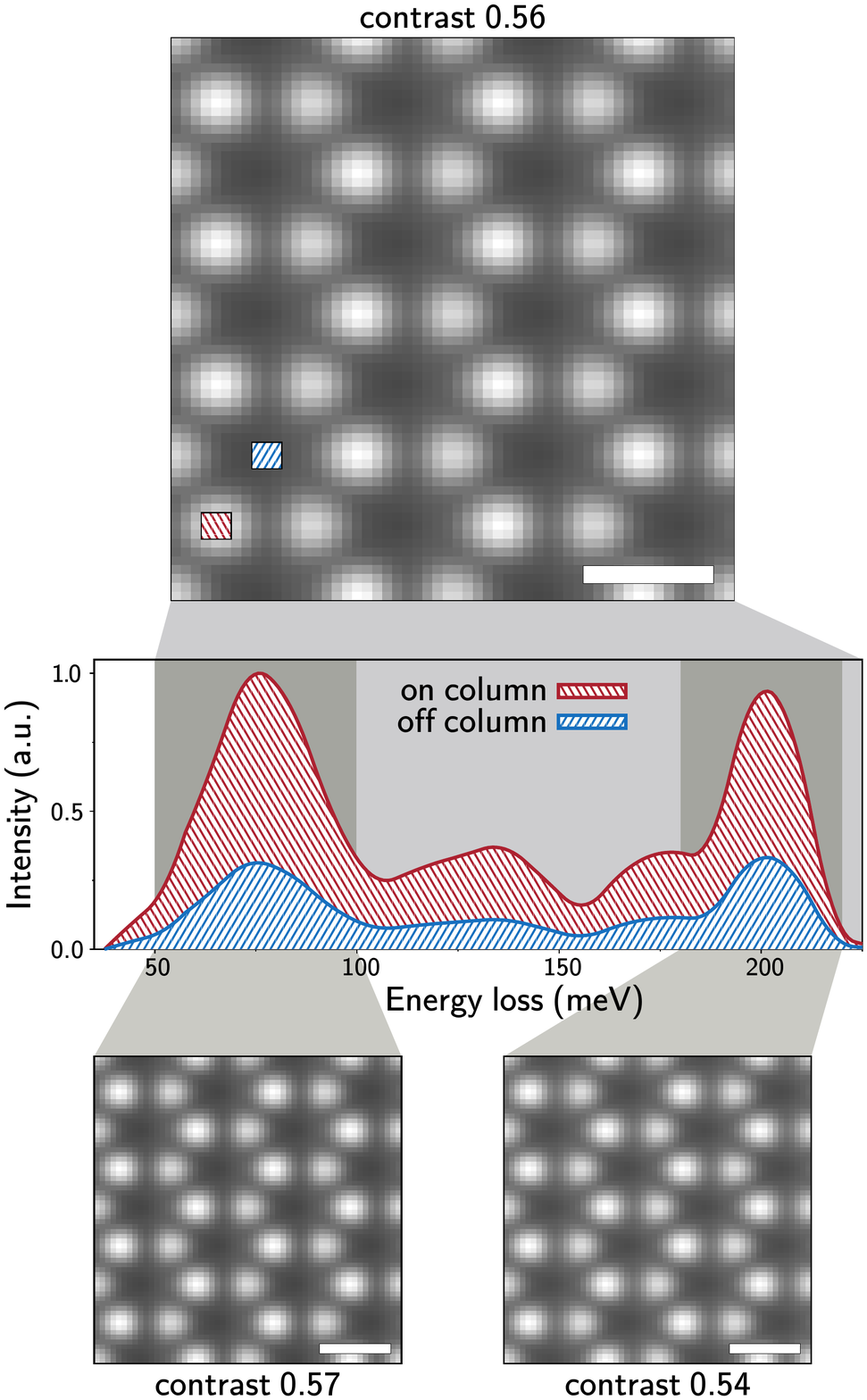}
    \caption{Dark field results: in the center the broadened vibrational electron energy loss spectra at the positions indicated in the top image are plotted. The top image is formed by integrating the corresponding spectrum over the energy range 50-225~meV at every pixel. The two lower images are formed by integrating over 50-100~meV and 180-220~meV. Refer to the main text for more details on the simulated conditions. The scale bars correspond to a distance of 2~\AA.}
    \label{fig:paperDF}
\end{figure}

We considered two different positions of the detector in our simulation, which are depicted in Fig. \ref{fig:method_schematics}. We refer to a detector, which is displaced by about 61~mrad along one of the Kikuchi bands in the diffraction pattern, a "dark field" detector and the bright field detector is centered with respect to the diffraction pattern. In both cases the detector covers a collection semiangle of 22~mrad.

Figure~\ref{fig:paperDF} displays an overview of the results for a dark field detector. The simulated spectra show clear peaks at around 75~meV and 200~meV, corresponding to scattering on acoustical and optical phonons, respectively. A clear difference in intensity between on and off column beam positions over the whole calculated energy range is observed in the spectra. By integrating the signal over energies between 50 and 225~meV according to equation \ref{eq:Ivib_image}, these intensity differences can be exploited to form an image resolving the positions of atomic columns, as shown at the top of Fig.~\ref{fig:paperDF} and revealing the hexagonal structure of $AA'$-stacked hBN. The atomic-scale contrast is furthermore preserved if the energy integration is performed over either the acoustical or optical phonon energies only. The contrast, defined as $(I_{\mathrm{max}} - I_{\mathrm{min}}) / (I_{\mathrm{max}} + I_{\mathrm{min}})$, is comparable for all three dark field images.

\begin{figure}
    \includegraphics[width=\linewidth]{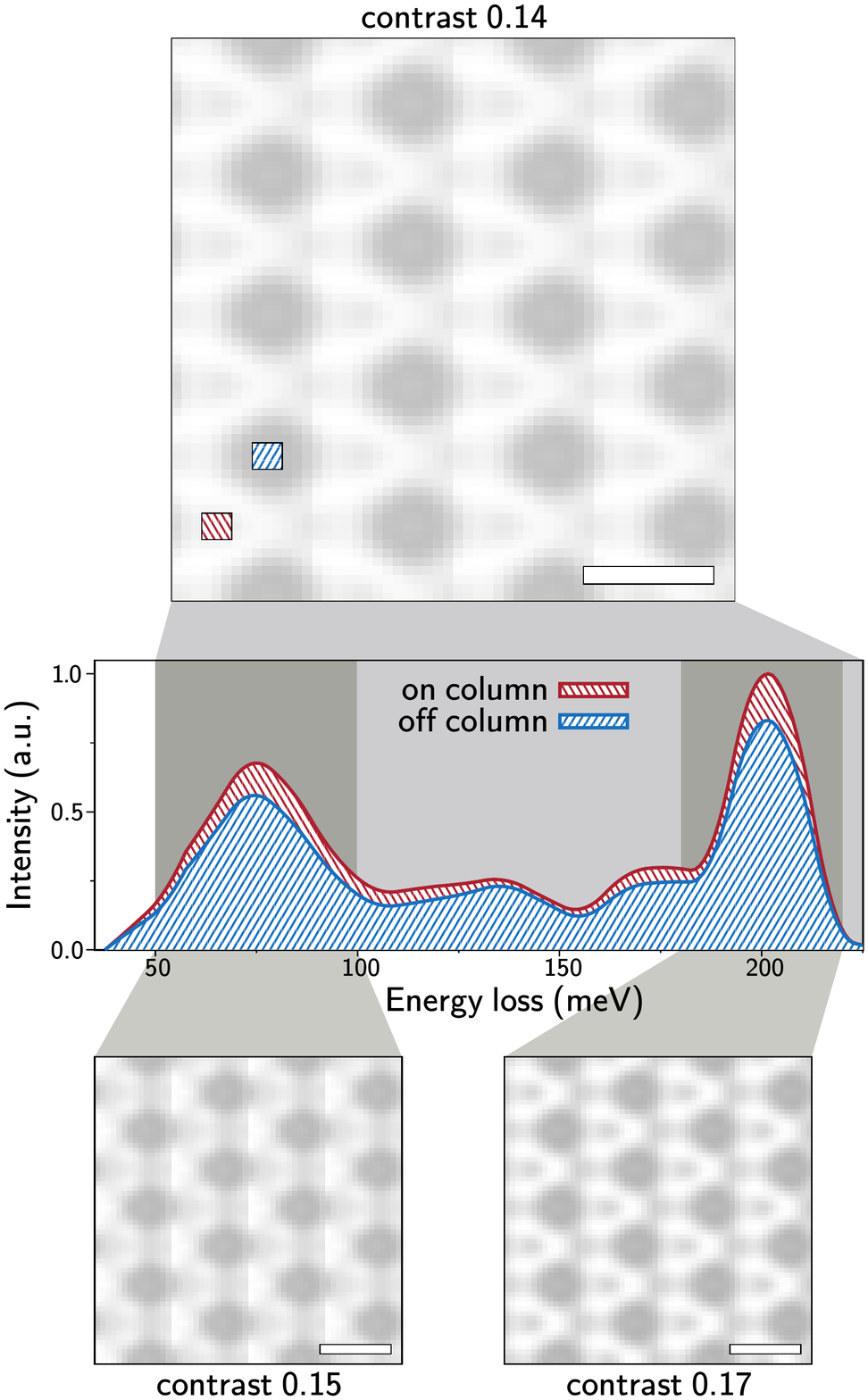}
    \caption{Bright field results: in the center, the broadened vibrational electron energy loss spectra at the positions indicated in the top image are plotted. The top image is formed by integrating the corresponding spectrum over the energy range 50-225~meV at every pixel. The two lower images are formed by integrating over 50-100~meV and 180-220~meV. Refer to the main text for more details on the simulated conditions. The scale bars correspond to a distance of 2~\AA.}
    \label{fig:paperBF}
\end{figure}

Figure~\ref{fig:paperBF} shows an overview of the results for a bright field detector. The simulated spectra show a much lower difference in intensity between on and off column beam positions than in the dark field case. This smaller difference leads to a much lower contrast in the image, which is formed by integrating over energies between 50~meV and 225~meV (at the top of Fig.~\ref{fig:paperBF}). For images, which take only energy ranges 50 to 100~meV and 180 to 220~meV into account, a much lower contrast than in their dark field counterparts is observed and atomic columns are not resolved. All bright field images exhibit, however, some vertical brightness streaks.

We proceed to compare our results with the experiment and simulation by Hage et al \cite{hage+19prl}. In the dark field, our images as well as the numerical value of the contrast agree very well with their images. In the bright field, the contrast value of the image for an energy range between 180 and 220~meV shown in Fig. \ref{fig:paperBF} agrees well with the contrast in the simulated bright field image presented by Hage et al.\ Both simulated images are qualitatively similar and show a "star of David"-like pattern, but each exhibits a different kind of artifact. Our image approaches the expected hexagonal symmetry, but displays the previously mentioned brightness streaks, whereas the hexagonal symmetry is mildly distorted in their image by the appearance of skew threads.

The appearance of brightness streaks in our images is an effect of our implementation of the frozen phonon simulation. Due to computational constraints, the same snapshot of the underlying vibrating crystal structure is used for all beam positions of the same energy bin. The difference between multislice passes for different beam positions is therefore only a real space shift of the optical axis of the beam, which leads to smooth images inside the calculated quarter of the rectangular conventional cell, but a sharp stripe contrast is present at the border.
For images averaged over 2 or 4 snapshots, these stripes are also visible in the dark field images but disappear for larger numbers. Overall the dark field images were visually well converged for 16 snapshots and 64 snapshots proved to be not sufficient to reach full convergence in the bright field. Analysis of the error in the intensity per pixel in the diffraction pattern shows, that it decreases much slower in the bright field than in the dark field case, since the relative strength of the vibrationally scattered intensity is much lower ($10^{-4}-10^{-6}$ vs $1-10^{-3}$). We could have chosen to forcibly symmetrize the image, a procedure mentioned by Loane et al.\ \cite{loan+91actacrystA47}, but we decided against such a treatment, in order to pinpoint  the difficulty of achieving convergence in the bright field.

We note a difference between simulated and the experimental spectra, namely that the peak corresponding to longitudinal optical and transverse optical (LO and TO) phonons is shifted by about 25~meV upwards in energy in our simulated spectra with respect to the measurement in Hage et al.\ The energy shift of the optical peak is an effect caused by the empirical interatomic potential we have used, which overestimates the energies of LO and TO phonon branches by around 10--40~meV along high symmetry directions in the Brillouin zone \cite{sevi+11prb}. We have therefore chosen the energy integration range for the image originating from optical phonons accordingly.

In terms of computational effort, %
the majority of the simulation time is spent on the multislice calculations, of which one requires $n_E \times n_{\mathrm{snap}}$ calculations per beam position. In this work we have performed $18 \times 64 = 1152$ multislice calculations per beam position, although for the dark field detector already $n_\mathrm{snap}=16$ snapshots were sufficient for well converged results, leading to 288 multislice calculations per beam position.

A general treatment of inelastic phonon scattering using a multislice solution to Yoshioka's equation is computationally very demanding \cite{wang89actacrystA45}. Under a single inelastic scattering approximation, 
it requires to compute the elastic propagation of the initial wave function to an atomic site, where the inelastic transition happens. After the inelastic interaction, the electron wave function is elastically propagated to the exit surface of the specimen. Since inelastic scattering from different atoms is to be treated incoherently, this procedure needs to be carried out for each atom in the simulation box and each mode. 
Typically, this requires about $n_{\textrm{modes}} \times n_{\mathrm{at}}$ multislice calculations, where $n_{\textrm{modes}} = 3 \times n_{\textrm{at}}$ is the number of considered phonon modes and $n_{\mathrm{at}}$ is the number of atoms in the simulation box.
Thus, for an equally sized supercell and associated phonon wave-vector grid, as used in this work, one would need to perform $n_{\textrm{modes}} \times n_{\mathrm{at}} = 3 \times n_{\mathrm{at}}^2 \approx 10^9$ multislice calculations per beam position, unless some reduced subset of phonon modes would be considered. 

The complexity of the method employed by Dwyer \cite{dwye17prb96} requires fewer multislice calculations than discussed above, since the inelastic scattering is accounted for by a projected M{\o}ller potential within each slice. This method requires $n_{\textrm{slices}} \times n_{\textrm{modes}}$ multislice calculations, where $n_{\textrm{slices}}$ is the number of slices. Even with greatly reduced number of phonon modes down to $n_\text{modes}=3\times 4\times 4\times 1$ and with $n_\textrm{slices}$ (at minimum) equal to number of atomic layers, i.e., 46 for our structure model, the required 2208 multislices per beam position remains a considerably larger computational effort than in our method.

Beyond the scope of this work, a detailed discussion of approximations inherent to our method remains, such as neglect of dipole scattering contributions \cite{dwye17prb96,rez14micromicroanal20}, treatment of multi-phonon and multiple phonon excitations \cite{wang98actacryst,wang89actacrystA45}, effects of anharmonicity in the interatomic potentials \cite{ceri-parr10pcs}, inaccuracies resulting from parametrized interatomic potentials and tractable ways to go beyond \cite{hung-cart09cpl475,shao+18cpc233,venk-ramp15intjqchem115}. All these aspects need attention and outline future directions stemming from this work.

We have demonstrated a frozen phonon method for the simulation of vibrational electron energy loss spectra for pure impact scattering. The method agrees with other published results and improves over other simulation techniques by combining the ability to simulate spectra with dynamical diffraction to all orders and the ability to simulate different scattering geometries. The method comes at reasonable computational effort and is straightforward to implement. It thus provides an efficient and versatile method for detailed simulations of vibrational spectroscopic experiments at high spatial resolution.

\begin{acknowledgments}
We acknowledge Venkat Kapil and Michele Ceriotti from EPFL in Lausanne for an introduction to i-Pi and valuable discussions. This research is funded by the Swedish Research Council and Swedish National Infrastructure for computing (SNIC) at the NSC center (cluster Tetralith).
\end{acknowledgments}

\bibliography{vib_STEM-EELS_arXiv}

\end{document}